\documentclass[pre,twocolumn,superscriptaddress,amsmath,amssymb,floatfix,showpacs]{revtex4}
\usepackage{graphicx}
\usepackage{dcolumn}
\usepackage{bm}
\begin{document}

\bibliographystyle{apsrev}

\title{Complete Wetting of Pits and Grooves}

\author{M. Tasinkevych}
\affiliation{Max-Planck-Institut f\"{u}r Metallforschung,
             Heisenbergstr. 3, D-70569 Stuttgart, Germany}
\affiliation{Institut f\"{u}r Theoretische und Angewandte Physik, 
         Universit\"{a}t Stuttgart, Pfaffenwaldring 57, 
         D-70569 Stuttgart, Germany}
\author{S. Dietrich}
\affiliation{Max-Planck-Institut f\"{u}r Metallforschung,
             Heisenbergstr. 3, D-70569 Stuttgart, Germany}
\affiliation{Institut f\"{u}r Theoretische und Angewandte Physik, 
         Universit\"{a}t Stuttgart, Pfaffenwaldring 57, 
         D-70569 Stuttgart, Germany}

\date{\today}

\begin{abstract}
For one-component volatile fluids governed by dispersion forces an effective interface Hamiltonian, derived from a microscopic 
density functional theory, is used to study complete wetting of geometrically structured substrates. Also the long range
of substrate potentials is explicitly taken into account.
Four types of geometrical patterns  are considered: (i) one-dimensional
periodic arrays of rectangular or parabolic grooves and (ii) two-dimensional lattices of
cylindrical or parabolic pits.  We present numerical evidence that at the centers of the
cavity regions the thicknesses of the adsorbed films  obey precisely the same geometrical covariance relation, 
which has been recently reported for complete cone and wedge filling. However, this covariance does not hold for the laterally averaged wetting
film thicknesses. For sufficiently deep cavities with vertical walls and close to  liquid-gas phase
coexistence in the bulk, the film thicknesses exhibit an effective planar scaling regime, which 
as function of undersaturation is characterized by a power law with the common
 critical exponent $-1/3$ as for a flat substrate, but with the amplitude depending on the geometrical
features. 
\end{abstract}

\maketitle

\section{Introduction}

It is well  known that a non-planar topography of a substrate modifies  its wetting properties significantly.
Accordingly, experimental studies of  complete wetting on substrates patterned by a two-dimensional lattice of
nanopits \cite{gang-exper_prl:05}, a one-dimensional array of wedges with finite depths \cite{bruschi-exper_prl:02}, and
arrays of microscopic nonlinear cusps or of semicircular channels \cite{Bruschi:06:1}
demonstrate the strong influence of nanocavities on the adsorption behavior  relative to that on flat substrates.
Theoretical studies of  adsorption in  infinitely deep \cite{reimer:99} generalized wedges \cite{parry_nature:00,parry_jcp:00}
predict that such substrates modify the  wetting exponents describing the divergence of the wetting film thickness upon approaching liquid-gas coexistence.  
%
Although  nanopatterning of surfaces may result in  drastic changes of their wettability,
recent  theoretical studies  have revealed surprising  hidden symmetries, or so-called covariances,
which  relate  local adsorption properties
for  various different substrate geometries.

\begin{figure}[]
\begin{center}
\includegraphics[width=0.65\linewidth]{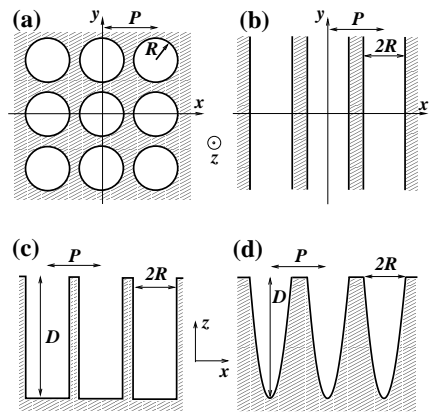}
\end{center}
\caption{ (a), (b) Top views of the considered substrate geometries. (a) Quadratic lattice with  lattice spacing $P$
 of  identical, cylindrical or parabolic  pits of  radius $R$. 
(b) Periodic array  of rectangular or parabolic grooves  of width $2R$. 
(c), (d) Cross sections along the plane $y=0$ of arrays of rectangular grooves or cylindrical pits (c), and
 parabolic grooves or pits (d).
All cavities have a finite depth $D$.
The grooves have a macroscopic extension in the $y$-direction.
}
\label{geometry}
\end{figure}
 An example of such a geometrical covariance relating  wedge and
cone complete filling  has been reported in Ref.~\cite{parry_prl:05}. It has been shown that the equilibrium midpoint
interfacial heights  $l^{(0)}$ in a {\it c}one and a {\it w}edge obey the relation 
$l_c^{(0)}(\Delta\mu,\alpha) = l_w^{(0)}(\Delta\mu /2,\alpha)$, where
$\alpha$ is the substrate tilt angle and the undersaturation $\Delta\mu\geq 0$ is  the chemical potential deviation 
from bulk liquid-gas coexistence.  This relation is valid for the leading asymptotic behaviors of $l^{(0)}_{c,w}$ in the limit 
 $\Delta\mu \rightarrow 0^+$.

In Ref.~\cite{tasinkevych:06:0} we have demonstrated that  complete wetting of substrates patterned by  periodic  arrays of grooves 
or quadratic lattices  of pits (see Fig.~\ref{geometry})  exhibits a geometrical covariance similar to the one 
described in Ref.~\cite{parry_prl:05}.  Moreover, we have also identified a range of undersaturations, within which
the midpoint interfacial height can be described by a {\em single} scaling function for all four types of substrate  patterns.
In this paper we extend the analyses of Ref.~\cite{tasinkevych:06:0}, showing inter alia
that,  in addition to the midpoint interfacial height, other parts of the interfacial profile  
exhibit the aforementioned covariance relation, too, while 
the laterally averaged interfacial profile does not. 

Our analysis is based on an effective interface Hamiltonian
approach \cite{dietrich_book,dietrich:91}. This approach has already been  applied earlier to study
wetting phenomena in a single groove \cite{darbellay_groove:92}, of a periodic array of grooves \cite{robbins:91}, 
in a wedge \cite{reimer:99,napiorkowski_wedge:92}, and also of chemically structured substrates \cite{koch:95,bauer_dietrich}.
Although this approach does not model the intrinsic structures of the emerging solid-liquid and 
of the depinning liquid-gas interfaces, it provides reliable results \cite{dietrich:91} concerning 
the growth and the morphology of  wetting films, phase transitions, etc. It has been shown 
that the leading asymptotic behavior of the effective interface potential does not depend
on the intrinsic  structures of the emerging solid-liquid and liquid-gas interfaces and is
given by the so-called sharp-kink approximation for the density distribution (see Sec.~\ref{model}) \cite{dietrich:91}.
Therefore, this effective interface model is a reliable approach for complete wetting considered here,
which probes the leading asymptotic behavior of the effective interface potential.

We consider rectangular or parabolic  grooves and cylindrical or parabolic pits, taking into account the  long
range of both the substrate potential and the fluid-fluid interaction. The temperature $T$ is chosen to be above the wetting
transition temperature $T_w$ of the unstructured planar substrate so that for $\Delta\mu \rightarrow 0^+$ complete wetting
occurs. (Similar considerations hold for binary liquid mixtures upon approaching their demixing transition from the
mixed phase.) 
As function of undersaturation  complete wetting of such substrates reveals four different scaling regimes \cite{tasinkevych:06:0}: 
filling, postfilling, effective planar, and planar,
with the three crossover values between these four neighboring regimes denoted as  $\Delta\mu_{fil}^{p,g}>\Delta\mu_{\pi}^{e}>\Delta\mu_{\pi}>0$.
Below we briefly describe all of them, providing also a short summary of the main results obtained.
The aforementioned covariance relates the behavior of the midpoint wetting film thicknesses  for all
geometries and holds within an undersaturation range $\Delta\mu_{\pi}^{e}\lesssim\Delta\mu\lesssim\Delta\mu_{fil}^{p,g}(R)$
(for $\Delta\mu_{\pi}^{e}$ see below, the superscripts $p,g$ refer to {\it p}its and {\it g}rooves, respectively) which we call
the  {\em postfilling scaling regime}.
 In the case of cylindrical pits or rectangular grooves, for $\Delta\mu\searrow \Delta\mu_{fil}^{p,g}(R)$ 
and for sufficiently large $D/R$, the analogue of  capillary condensation occurs such that the pits or grooves are
 rapidly, but continuously, filled by the liquid.  However, in the case of the parabolic pits and grooves,
$\Delta\mu_{fil}^{p,g}$ marks the crossover from the continuous power-law filling for $\Delta\mu>\Delta\mu_{fil}^{p,g}$ to the
postfilling scaling regime. 
We find numerically for  $R/\sigma\gtrsim 50$ (see Fig.~\ref{geometry}), where $\sigma$ is a molecular length scale,  
 $\Delta\mu_{fil}^{p,g}(R)\sim R^{-1-\delta}$ with a small positive effective
exponent $\delta$, and $\Delta\mu_{fil}^{p} =2\Delta\mu_{fil}^{g}$. 

If we denote the equilibrium interface height
at the position of the symmetry axes of the cylindrical or parabolic pits as $l_p^{(0)}$, and in the middle of the rectangular or parabolic
 grooves as $l_g^{(0)}$,  we obtain in the postfilling regime
\begin{equation}
l_{p,g}^{(0)}(\Delta\mu,R,P,D)=R\Lambda_{p,g}\biggl ( \frac{\Delta\mu}{\varepsilon_f} \biggl(\frac{R}{\sigma} \biggr)^{1+\delta} \biggr )
\label{eq:main_1a}
\end{equation}
and
\begin{equation}
\Lambda_p(v)=\Lambda_g(v/2),
\label{eq:main_1b}
\end{equation}
where $\varepsilon_f$ is a molecular energy scale.
Equation (\ref{eq:main_1b}) expresses the same covariance relation as the one reported in 
Ref.~\cite{parry_prl:05}.
 The dimensionless scaling functions $\Lambda_{p,g}(x)$ of the postfilling regime 
 do not depend on  $D$  or $P$, i.e., within this regime the midpoint interfacial heights 
increase upon decreasing undersaturation in the same way for an isolated cavity as for arrays of them.

For $\Delta\mu<\Delta\mu_{\pi}^{e}$ the  cavities are completely filled by the liquid and the equilibrium local interface 
heights $l_{p,g}(x,y)=l_{p,g}$  become de facto independent of the lateral coordinates ${\bf x}\equiv (x,y)$. 
In the case of the rectangular grooves or cylindrical pits, 
$\Delta\mu_{\pi}^{e}$ marks the crossover from the postfilling scaling regime
to the {\em effective planar scaling regime} within which the wetting behavior of geometrically patterned substrates can be  mapped 
onto that of  layered, laterally homogeneous solids.
 The upper layer of those ersatz solids has a thickness $D$ and its composition is related 
to the geometrical parameters  $R$ and $P$. This results in the following scaling relations for the
effective planar scaling regime:
\begin{eqnarray}
l_{p,g}(\Delta\mu,R,P,D)&=&(\Phi_{p,g})^{\frac{1}{3}}l_{\pi}(\Delta\mu),
 \nonumber \\
\Phi_{p}\equiv 1-\pi\biggl (\frac{R}{P}\biggr )^2&,& \Phi_{g}\equiv 1-\frac{2R}{P},
\label{eq:gc_planar_H}
\end{eqnarray} 
where $l_{\pi}$ is the thickness of the adsorbed liquid film on the corresponding original non-structured planar substrate. 
$\Phi_{p}$ and $\Phi_{g}$ are the areal fractions of solid in the top layer, $z=0$, of substrates
with cylindrical pit and rectangular groove patterns, respectively.
 For $D\rightarrow 0$ the width of the effective planar regime, i.e., the range 
of applicability of Eq.~\ref{eq:gc_planar_H} vanishes. 


 Finally, at $\Delta\mu=\Delta\mu_{\pi}\sim D^{-3}$ with $\Delta\mu_{\pi}<\Delta\mu_{\pi}^{e}$ the systems 
cross over to the {\em planar scaling} regime, in which the geometrical patterns are irrelevant.
In the case of the parabolic pits and grooves, we do not observe the effective planar scaling regime.  There is rather 
an extended crossover region from the postfilling scaling regime to the planar one. 

In Sect.~\ref{model} we provide 
connections between  microscopic density functional theory and the effective interface potential 
approach. In Sect.~\ref{results} we describe in detail the four scaling regimes, which 
are revealed by the midpoint interfacial heights as functions of undersaturation.
Sect.~\ref{theory_vs_experiment}  provides a comparison with experimental data and  Sect.~\ref{summary} summarizes our main results and provides an outlook.

\section{Effective Interface Hamiltonian}
\label{model}

We start from the following  grand canonical density functional for 
inhomogeneous fluids characterized by the number density $\rho({\bf r})$ \cite{evans:79}:
\begin{eqnarray}
 \Omega[\{\rho({\bf r})\},T,\mu] &=& \int d^3r f_{HS}(\rho({\bf r}),T) \nonumber \\ 
&+&\frac{1}{2} \int d^3r\int d^3r^{\prime}w(|{\bf r}-{\bf r^{\prime}}|)\rho({\bf r}) \rho({\bf r^{\prime}})\nonumber \\
 &+&\int d^3r(\rho_s V({\bf r})-\mu)\rho({\bf r}). 
\label{DFT}
  \end{eqnarray}
The integrations are performed over the region accessible to the fluid; $w(r)$ describes the long-ranged attractive part of the fluid-fluid
interaction potential and decays at large distances $\sim r^{-6}$. The short-ranged repulsion is treated according to the 
Weeks-Chandler-Andersen (WCA) approximation \cite{WCA}  by introducing a reference system of hard spheres with bulk 
free energy density $f_{HS}(\rho,T)$; $\mu$ is the chemical potential and $V({\bf r})$ is the
substrate potential, which can be obtained by integrating the fluid-{\it s}ubstrate pair potential $w_{s}(r)$ over
the region occupied by the substrate particles with  number density $\rho_s$. 
The equilibrium density distribution minimizes $\Omega$. In order to make analytic progress we seek the minimum
of $\Omega[\rho({\bf r})]$  within the subspace of steplike varying density profiles \cite{dietrich_book,dietrich:91}:
\begin{eqnarray}
\rho({\bf r}=({\bf x},z))&=&\Theta\Bigl( z-s({\bf x}=(x,y))\Bigr ) \nonumber \\
&\times&\Bigl (\rho_l \Theta(l({\bf x}) -z)+\rho_g \Theta(z- l({\bf x})) \Bigr),
\label{kink}
\end{eqnarray}
where $\rho_l$ and $\rho_g$ are the number densities of the bulk liquid and gas phases, respectively;
 $s({\bf x})$ denotes the substrate surface which is in contact with the fluid;
$z=l({\bf x})$ is the local position of the liquid-gas interface assuming the absence of bubbles and 
overhangs; and $\Theta$ is the Heaviside step function.
 Inserting Eq.~(\ref{kink}) into Eq.~(\ref{DFT}) yields the grand canonical potential
$\Omega$ as a functional of the interface morphology $l({\bf x})$:
\begin{eqnarray}
\Omega[\{l({\bf x})\};\rho_l,\rho_g,\mu] &=& {\it const} \nonumber \\  + \int\limits_{{\cal A}}d^2x\Bigl ( \Delta\Omega l({\bf x}) 
&+&\Sigma_{lg}({\bf x},l({\bf x})) +  W({\bf x},l({\bf x})) \Bigr).
\label{eff_hamilton_non_local}
\end{eqnarray}
${\cal A}$ is a macroscopic area, ${\it const}$ denotes the sum of all terms independent of $l({\bf x})$, and  $\Delta\Omega$ 
is the difference of the grand
canonical potential densities of the uniform bulk liquid and gas phases, i.e.,
 $\Delta\Omega\equiv \Omega_b(\rho_l,T,\mu) - \Omega_b(\rho_g,T,\mu)$. Close to the bulk liquid-gas coexistence line $\mu=\mu_0(T)$
 one has $\Delta\Omega=\Delta\rho\Delta\mu+O(\Delta\mu^2)$, with $\Delta\mu=\mu_0(T)-\mu\leq 0$, and  $\Delta\rho=\rho_l-\rho_g$.
$\Sigma_{lg}({\bf x},l({\bf x}))$ describes the cost in free energy to maintain a laterally inhomogeneous
interface configuration relative to that for the flat reference configuration (see Ref.~\cite{napiorkowski:91:0}):
\begin{eqnarray}
&&\Sigma_{lg}({\bf x},l({\bf x}))= \nonumber \\
&&-(\Delta\rho)^2\int\limits_{{\cal A}}d^2x^{\prime}\int\limits_0^{\infty}dz\int\limits_0^{l({\bf x})-l({\bf x^{\prime}})}dz^{^{\prime}}w(|{\bf r-r^{\prime}}|).
\label{sigma_non_local}
\end{eqnarray}
The functional form of the effective interface potential $W({\bf x},l({\bf x}))$  depends on the substrate shape as well as
the  fluid-fluid and the substrate-fluid interaction potentials:
\begin{eqnarray}
W({\bf x},l({\bf x})) = \Delta\rho \int\limits_{{\cal A}}d^2x^{\prime} 
\int\limits_{l({\bf x})-s({\bf x^{\prime}})}^{\infty} dz \int\limits_z^{\infty}dz^{\prime} \nonumber \\ 
\times \Bigl ( \rho_l w(\sqrt{({\bf x-x^{\prime}})^2+{z^{\prime}}^2})-\rho_s w_{s}(\sqrt{({\bf x-x^{\prime}})^2+{z^{\prime}}^2})\Bigr ).
\label{interface_potential}
\end{eqnarray}

\begin{figure}
\begin{center}
\includegraphics[width=0.8\linewidth]{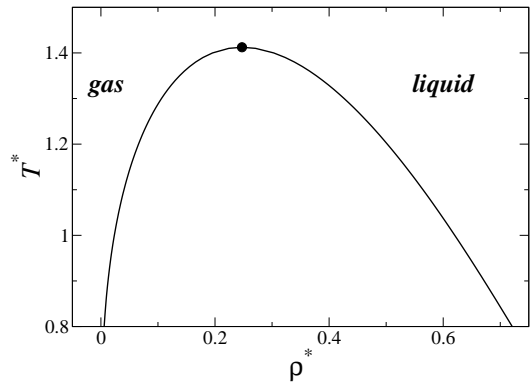}
\end{center}
\caption{Liquid-gas bulk phase diagram in terms of $T^*=k_BT/\varepsilon_f$ and $\rho^*=\rho\sigma^3$.
The critical point is given by $(T^*_c,\rho^*_c)\approx(1.4124,0.2473)$.
}
\label{bulk_diagram}
\end{figure}
\noindent The fluid particles are assumed to interact via a Lennard-Jones potential
\begin{equation}
\phi (r) = 4 \varepsilon_f \biggl [ \Bigl (\frac{\sigma}{r} \Bigr )^{12} - \Bigl ( \frac{\sigma}{r} \Bigr )^{6} \biggr ].
\end{equation}
We apply the WCA \cite{WCA} procedure to split  up the interaction into a repulsive part $\phi_{rep}(r)$ and an attractive part
$\phi_{att}(r)$, where
\begin{equation}
 \phi_{rep}(r) =
\begin{cases}
\phi (r) + \varepsilon_f, & ~~ r<2^{1/6}\sigma \\
0,                        & ~~ r\geq 2^{1/6}\sigma  \\
\end{cases}
\label{phi_rep}
\end{equation}
and
\begin{equation}
 \phi_{att}(r) =
\begin{cases}
 -\varepsilon_f, & ~~ r<2^{1/6}\sigma \\
\phi (r),        & ~~ r\geq 2^{1/6}\sigma.  \\
\end{cases}
\label{phi_att}
\end{equation}
The repulsive part gives rise to an effective, temperature dependent, hard sphere diameter 
\begin{equation}
d(T) = \int\limits_0^{2^{1/6}\sigma}dr\biggl [ 1-{\rm exp}\Bigl (-\frac{\phi_{rep}(r) }{k_BT} \Bigr ) \biggr ].
\end{equation} 
\noindent This expression is inserted into the Carnahan-Starling approximation for the free energy density of the hard-sphere fluid \cite{carnahan:69}
\begin{eqnarray}
f_{HS}(\rho,T)&=&k_BT\rho\biggl ({\rm ln}(\rho\lambda^3)-1+\frac{4\eta-3\eta^3}{(1-\eta)^2} \biggr ), \\ \nonumber
\eta &=& \frac{1}{6}\pi\rho d(T)^3,
\end{eqnarray}
where $\lambda$ is the thermal de Broglie wavelength.
The attractive part $\phi_{att}(r)$ of the fluid-fluid interaction potential is approximated by
\begin{equation}
w(r)   = w_0\frac{4\sigma^3}{\pi^2}(\sigma^2+r^2)^{-3},
\label{ff_potential}
\end{equation}
such that the integrated strength of $w(r)$ equals that of the original $\phi_{att}(r)$:
\begin{equation}
w_0 = \int\limits_{{\mathbb R}^3}d^3r w(r) = \int\limits_{{\mathbb R}^3}d^3r \phi_{att}(r) = -\frac{32}{9}\sqrt{2} \pi \varepsilon_{f}\sigma^3.
\end{equation}
Equation (\ref{ff_potential}) resembles, but differs from the WCA expression $\phi_{att}(r)$  for the attractive potential; 
this approximation offers the benefit to be able to perform certain
integrations analytically. 
\noindent The actual values $\rho_l$ and $\rho_g$ minimize $\Omega_b(\rho,T,\mu)$, and the coexistence line
$\mu=\mu_0(T)$ is determined by the following equations:
\begin{equation}
\label{bulk_densities}
\left.{\frac{\partial\Omega_b}{\partial\rho}}\right\vert_{\rho=\rho_l} = \left.{\frac{\partial\Omega_b}{\partial\rho}}\right\vert_{\rho=\rho_g} = 0
\end{equation}
\noindent and
\begin{equation}
\Omega_b(\rho_l) = \Omega_b(\rho_g).
\label{equlibrium_cond}
 \end{equation}
The grand canonical potential density of the bulk system has the form
\begin{equation}
\Omega_b(\rho,T,\mu) = f_{HS}(\rho,T) + \frac{1}{2}w_0\rho^2-\mu\rho.
\label{bulk_grand_pot}
\end{equation}
The bulk phase diagram resulting from Eqs.~(\ref{bulk_densities})-(\ref{bulk_grand_pot}) is shown in Fig.~\ref{bulk_diagram}.

Since here we are only  interested in complete wetting involving sufficiently thick films, the  functional form in Eq.~(\ref{ff_potential}) can also
be used to model the interaction potential between fluid and substrate particles, because under
these circumstances only its asymptotic behavior at large distances matters:
 \begin{equation}
w_s(r)  = -\frac{128\sqrt{2}}{9\pi}\varepsilon_{s}\sigma_s^6(\sigma_s^2+r^2)^{-3}.
\label{sf_potential}
\end{equation}
In the following we use $\sigma_s = \sigma$.
This leads to the following expression for the effective interface potential:
\begin{eqnarray}
W({\bf x},l({\bf x}))&=&A\times I_{V_s}({\bf x},l({\bf x})), \nonumber \\
I_{V_s}({\bf x},l({\bf x})) &=& \int\limits_{l({\bf x})}^{\infty}dz\int\limits_{V_s}d^3r^{\prime}(\sigma^2+|{\bf r}-{\bf r^{\prime}}|^2)^{-3},
\label{interface_potential-2}
\end{eqnarray}
 where $A=-\frac{128\sqrt{2}}{9\pi}\sigma^6\Delta\rho(\rho_l\varepsilon_{f}-\rho_s\varepsilon_{s})>0$ plays the role of
an effective  Hamaker constant, and $V_s$ denotes the 
domain occupied by  the substrate particles which are assumed to be homogeneously distributed within
$V_s$ with number density $\rho_s$.
The effective interface potential of  the planar substrate occupying the region $z\leq 0$ is
\begin{equation}
W_{\pi}(l) = \frac{A\pi}{4\sigma^3}\Bigl ( \sigma -\frac{\pi l}{2}+l\arctan(\frac{l}{\sigma}) \Bigl ),
\label{interface_potential_planar}
\end{equation}
with $W_{\pi}(l\gg\sigma)\approx A\pi/(12l^2)$ so that $l_{\pi}(\Delta\mu\rightarrow 0)=(A\pi/(6\Delta\rho\Delta\mu))^{1/3}$.
By applying a small gradient expansion \cite{napiorkowski:91:0} for the nonlocal expression  $\Sigma_{lg}({\bf x},l({\bf x}))$ 
one obtains the familiar local approximation for  the grand canonical potential (see Eq.~(\ref{eff_hamilton_non_local})):
\begin{eqnarray}
\Omega^{loc}[\{l({\bf x})\};\rho_l,\rho_g,\mu]       &=& const \nonumber \\ 
+ \int d^2x \Bigl  (\sigma_{lg}\sqrt{1+(\nabla l)^2} &+&
\Delta{\mu} \Delta\rho l + W({\bf x},l) \Bigr );
\label{eff_hamilton_local}
\end{eqnarray}
$\sigma_{lg}$ is the surface tension of the free liquid-vapor interface, which  within the present sharp-kink approximation
has the following form:
\begin{eqnarray}
\sigma_{lg} &=& -\frac{1}{2}(\Delta\rho)^2
\int\limits_0^{\infty}dz\int\limits_z^{\infty}dz^{\prime}\int\limits_{\cal A}d^2x w(\sqrt{{\bf x}^2 + z^{\prime 2}}) \nonumber \\
&=& \frac{16\sqrt{2}(\Delta\rho\sigma^3)^2 \varepsilon_f}{9\sigma^2}.
\label{surface_tension}
\end{eqnarray}

Having determined the bulk coexistence curve $\mu_0(T)$, in the following 
 we minimize the functional in Eq.~(\ref{eff_hamilton_local}) numerically  which yields the equilibrium interface height
$l({\bf x})$ within mean-field approximation.  
The bulk liquid and gas densities $\rho_l$ and
$\rho_g$, respectively, are evaluated at  coexistence, i.e., $\rho_l=\rho_l(T,\mu_0(T))$ and
$\rho_g=\rho_g(T,\mu_0(T))$. 
All results presented below are obtained for $T/T_c=0.85$.
For the substrate structured with grooves the four integrals 
in the Eq.~(\ref{interface_potential-2}) can be calculated
analytically, while for the  two-dimensional lattices of cylindrical or parabolic pits
only three of them can be carried out analytically.
Whereas the capillary wave spectrum of $\Omega$ given by Eq.~(\ref{eff_hamilton_non_local})
differs qualitatively \cite{napiorkowski:91:0,mecke:99} from that of the local approximation
in Eq.~(\ref{eff_hamilton_local}), the morphology of the wetting film is captured reliably by 
Eq.~(\ref{eff_hamilton_local}) apart from  regions of high curvature \cite{bauer:99}
which do not play an important role in the present context. Therefore, it is legitimate to apply 
Eq.~(\ref{eff_hamilton_local})  even in the presence of long-ranged fluid-fluid
interactions, which are additionally encoded in $W({\bf x},l)$.

\section{Numerical Results}
\label{results}

\begin{figure}[]
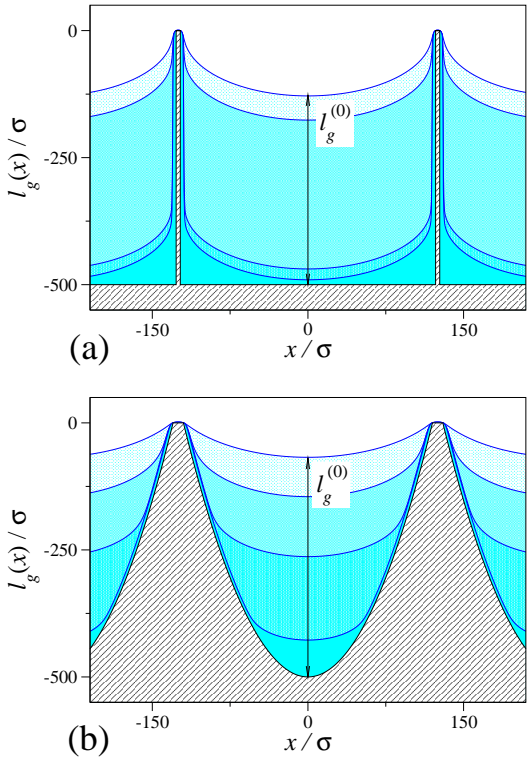

\begin{center}
\includegraphics[width=0.8\linewidth]{fig3a.eps}
\end{center}
\begin{center}
\includegraphics[width=0.8\linewidth]{fig3b.eps}
\end{center}
\caption{ Interfacial profiles $l_g(x)$ (a) for  rectangular grooves (which are translationally invariant 
in the $y$ direction)  and four values
of undersaturation (top to bottom): $\Delta\mu/\varepsilon_f = 0.00921,$ $0.00928,$ $0.00929,$ $0.0939$ which are  close to
$\Delta\mu_{fil}^g$, $R/\sigma =123$; (b) for 
 parabolic grooves  for $\Delta\mu/\varepsilon_f = 0.008,$ $0.01,$ $0.012,$ $0.02$ (top to bottom), $R/\sigma =120$.
For both structures $P/\sigma = 250$ and $D/\sigma = 500$ (see Fig.~\ref{geometry}). $l=0$ corresponds to
the top of the substrate structure at $z=0$. Here the effective Hamaker constant is $A/\varepsilon_f \approx  0.330$, and
the surface tension of the free liquid-gas interface is $\sigma_{lg}\sigma^2/\varepsilon_f\approx 0.478$.}
\label{profiles}
\end{figure}

 For grooves, in Fig.~\ref{profiles} typical
 interfacial profiles $l_g(x)$ are shown as obtained  for several values of $\Delta\mu$.
 We are primarily concerned with the behavior of the midpoint heights
 $l_{p,g}^{(0)}$ as functions of the undersaturation $\Delta\mu$. 
 For all substrate geometries the midpoint heights $l_{p,g}^{(0)}(\Delta\mu)$ exhibit four different regimes.

\subsection{Large $\Delta\mu$: filling and postfilling regimes}
\label{results-large-delta-mu}

\begin{figure}[]
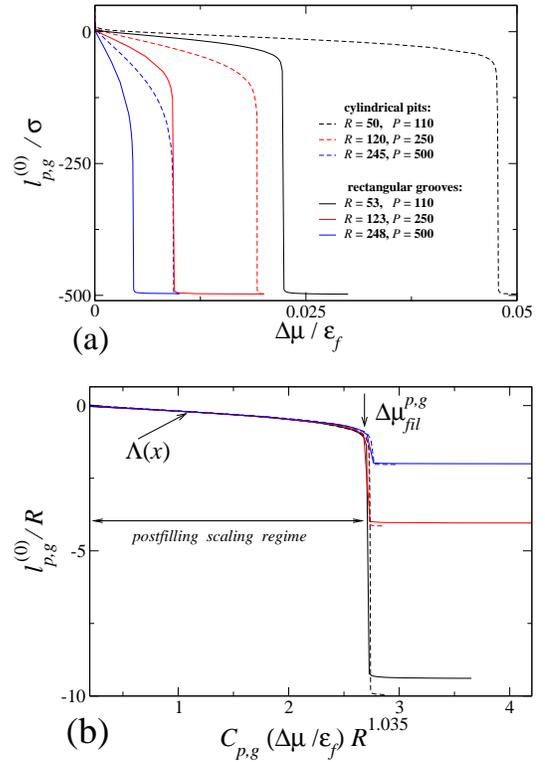

\begin{center}
\includegraphics[width=0.8\linewidth]{fig4a.eps}
\end{center}
\begin{center}
\includegraphics[width=0.8\linewidth]{fig4b.eps}
\end{center}
\caption{ (a) Interfacial heights  (measured from the top substrate layer at $z=0$) at the middle of the
cylindrical  pits, $l_p^{(0)}$ (dashed lines), and of the
rectangular grooves, $l_g^{(0)}$  (solid lines), as a function of 
undersaturation  $\Delta\mu$  measured in units of the energy scale $\varepsilon_f$ of the
fluid-fluid interaction.
(b) Rescaling the variables  according to Eqs.~(\ref{eq:main_1a}) and (\ref{eq:main_1b}) leads to data collapse
in the postfilling scaling regime; $C_p=1$ and $C_g=2$.
The depth of the cavities is $D=500$.
All length are measured in units of $\sigma$. 
$A/\varepsilon_f \approx  0.330$, and $\sigma_{lg}\sigma^2/\varepsilon_f\approx 0.478$.
$\Lambda (x)$ denotes the scaling function introduced 
in Eq.~(\ref{eq:main_1b}).}
\label{l0_rect}
\end{figure}

The first regime corresponds to the filling of the cavities. 
Cavities with vertical walls (rectangular grooves or cylindrical pits)
 are filled in a qualitatively different way compared to parabolic ones.
For the case of cylindrical {\it p}its and rectangular {\it g}rooves
and for $D/R$ large enough, there exists a well defined value of the undersaturation
$\Delta\mu=\Delta\mu_{fil}^{p,g}$ at which  an abrupt, but still continuous, filling of 
the  cavities takes place.  A similar behavior has been reported earlier for an isolated  
rectangular groove \cite{darbellay_groove:92}. The filling of the vertical cavities is shown in Fig.~\ref{l0_rect}(a),
 where $l_{p,g}^{(0)}(\Delta\mu)$ are shown for several values of $R$ and $P$. 
The interfacial profiles, which are shown in Fig.~\ref{profiles}(a), are calculated for undersaturations close to
 $\Delta\mu_{fil}^{g}$,  $\vert \Delta\mu - \Delta\mu_{fil}^{g}\vert < 0.002$, 
and  $R/\sigma = 123$.
$\Delta\mu_{fil}^{p,g}$  depend  only on the lateral size  $R$ of the  cavities, but not on the depth $D$.
We find numerically that  $\Delta\mu_{fil}^{p,g} \sim  R^{-1-\delta}$, with an effective exponent $\delta\approx 0.035$ for both 
the cylindrical pits and the rectangular grooves.

\begin{figure}[]
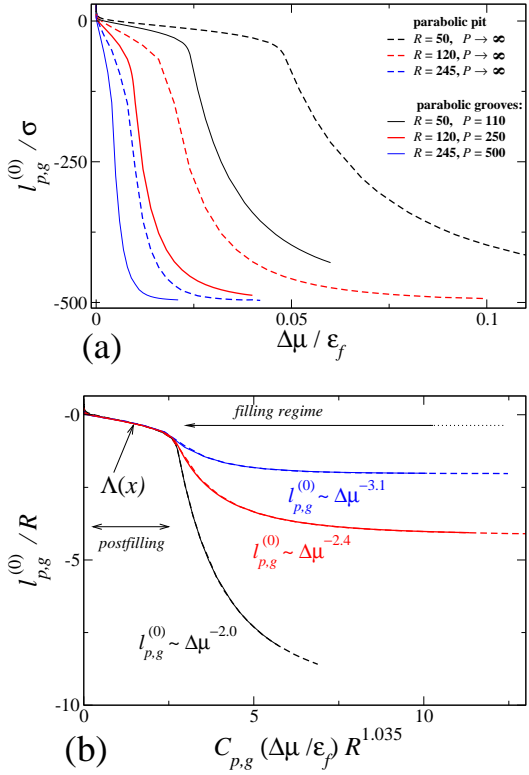

\begin{center}
\includegraphics[width=0.8\linewidth]{fig5a.eps}
\end{center}
\begin{center}
\includegraphics[width=0.8\linewidth]{fig5b.eps}
\end{center}
\caption{ (a) Interfacial heights  at the middle of the
parabolic  pit, $l_p^{(0)}$ (dashed lines), and of the
parabolic grooves, $l_g^{(0)}$  (solid lines), as a function of  $\Delta\mu$.
(b) Rescaling the variables  according to Eqs.~(\ref{eq:main_1a}) and (\ref{eq:main_1b}) leads to data collapse
 for $\Delta\mu\in (\Delta\mu_{\pi}^e,\Delta\mu_{fil}^{p,g})$; $C_p=1$ and $C_g=2$.
The depth of the cavities is $D=500$.
All length are measured in units of $\sigma$. 
$A/\varepsilon_f \approx  0.330$, and $\sigma_{lg}\sigma^2/\varepsilon_f\approx 0.478$.
$\Lambda (x)$ denotes the scaling function introduced 
in Eq.~(\ref{eq:main_1b}). The filling regime ends at large $\Delta\mu$ (dotted line)
where the wetting films in the center become microscopicly thin so that scaling breaks down.}
\label{l0_parab}
\end{figure}

Different from the previous case, the complete filling of the parabolic cavities  is described by an
effective power law, $l_{p,g}^{(0)}(\Delta\mu,R,D)\sim\Delta\mu^{-\gamma(R,D)}$, which is 
valid for $\Delta\mu \gtrsim \Delta\mu_{fil}^{p,g}$ only (see Fig.~\ref{l0_parab}(a)).
The effective exponent $\gamma$ ranges  from ca. $ 3.1$, for $R=245\sigma$, to ca. $2.0$, for $R=50\sigma$, 
at a cavity depth $D=500\sigma$. The value $\gamma = 2$ is expected to hold  for an infinitely deep, single
 parabolic wedge \cite{parry_nature:00}. For both vertical and parabolic cavities, the complete filling  obeys the covariance relation
$l_p^{(0)}(\Delta\mu,R,D)=l_g^{(0)}(\Delta\mu/2,R,D)$ valid  for $\Delta\mu \gtrsim \Delta\mu_{fil}^{p,g}$.

For  $\Delta\mu <\Delta\mu_{fil}^{p,g}$ the midpoint height
for all  patterns shows an almost linear dependence on $\Delta\mu$ on normal scales.
The interfacial profiles still reflect the geometrical patterns, i.e., there are  
considerable lateral variations of the interfacial heights. For large enough $D$ there exists
a range of undersaturations  $(\Delta\mu_{\pi}^e,\Delta\mu_{fil}^{p,g})$ within which the midpoint
interfacial height for all four substrate geometries can be described by a {\em single} scaling function \cite{tasinkevych:06:0}. We call this
range of    $\Delta\mu$ the postfilling scaling regime. The lower bound $\Delta\mu_{\pi}^e$ of this regime 
is a decreasing function of the cavities size $R$.
We find that  for $\Delta\mu\in (\Delta\mu_{\pi}^e,\Delta\mu_{fil}^{p,g})$ the slopes of the $l_{p,g}^{(0)}$ 
curves scale as $R^\alpha$ with $\alpha\approx 2$. Combining  this result with the scaling of the chemical potential 
$\Delta\mu_{fil}^{p,g}$ at the crossover to the filling scaling regime,
we propose for the functions $l_{p,g}^{(0)}$ in the postfilling regime the scaling forms  given by Eq.~(\ref{eq:main_1a}).
In this postfilling scaling regime the scaling function $\Lambda_{g}$ describes  rectangular as well as parabolic grooves, and similarly  
$\Lambda_{p}$ describes cylindrical and parabolic pits. 
\begin{figure}[]
\begin{center}
\includegraphics[width=0.8\linewidth]{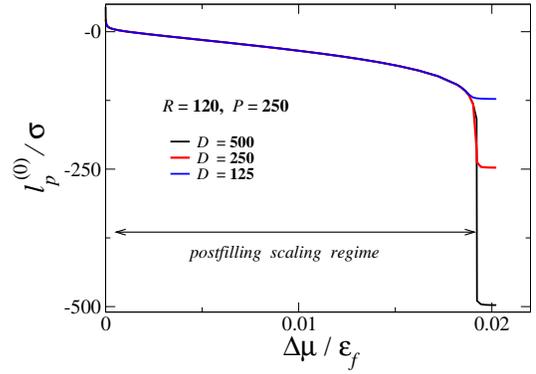}
\end{center}
\caption{Interfacial heights $l_p^{(0)}$ in the middle of the
cylindrical  pits   as a function of 
undersaturation  $\Delta\mu$ for several values of the pit depth $D$ (see Fig.~\ref{geometry}).
All lengths are measured in units of $\sigma$;
$A/\varepsilon_f \approx  0.330$ and $\sigma_{lg}\sigma^2/\varepsilon_f\approx 0.478$.}
\label{l0_cyl_eff_depth}
\end{figure}

\noindent The scaling functions $\Lambda_{p,g}$ and the corresponding data collapse upon suitably rescaling
 the chemical  potential and the midpoint heights are shown in  Figs.~\ref{l0_rect}(b) and 5(b) for the case of
vertical and parabolic cavities, respectively.
In accordance with Eq.~\ref{eq:main_1b} the difference between $\Lambda_p(x)$ and $\Lambda_g(x)$ can be absorbed
in the rescaling factors $C_p=1$ and $C_g=2$. Inspection tells that the resulting common
scaling function denoted as $\Lambda(x)$  in Figs.~\ref{l0_rect}(b) and \ref{l0_parab}(b) are indeed identical.
The scaling functions  $\Lambda_{p,g}$ in the postfilling scaling regime do not depend on the cavity depth $D$ which is 
demonstrated in Fig.~\ref{l0_cyl_eff_depth},  where $l_{p}^{(0)}(\Delta\mu)$ is shown in the case of cylindrical pits and for
several values of $D$.   For even smaller  values of $D/R$, owing
to the finite-size rounding effects, the filling of the cavities does not occur at a certain value of the chemical potential, 
but there is an entire transition region. As a result, the range of the postfilling scaling regime 
shrinks as $D/R$ decreases.
 $\Lambda_{p,g}$ do not depend on the pattern
periodicity $P$ either. In the case of a single cavity, i.e., in the limit $P\rightarrow\infty$, we obtain 
 the same curves $l_{p,g}^{(0)}(\Delta\mu)$  as those  presented in  Figs.~\ref{l0_rect} and \ref{l0_parab}. 
Furthermore, we find that for the filling and postfilling scaling regimes not only $l_{p,g}^{(0)}$,  but
also the full interfacial profiles $l_{p,g}({\bf x})$ are,
within numerical accuracy,  independent of  the pattern periodicities $P$ studied here. This implies that for the case
of pits (both cylindrical and parabolic) the interfacial profile within the cavity region is, to a
good approximation,  rotational symmetric around the symmetry axis of the pit, i.e.,
$l_p({\bf x}) = l_p(r_p)$, where $r_p\lesssim R$ is the radial distance from the symmetry axis of a pit.

\begin{figure}[]
\begin{center}
\includegraphics[width=0.8\linewidth]{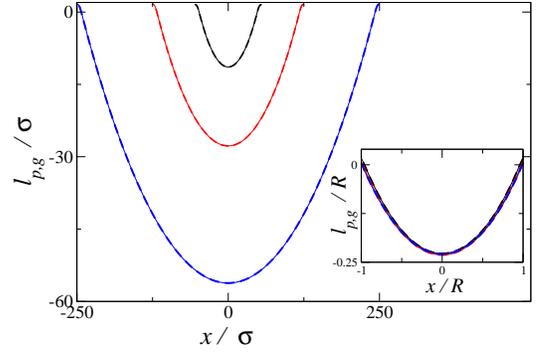}
\end{center}
\caption{Interfacial profiles within the cavity regions for cylindrical pits, $l_p(x,y=0)$ (dashed lines), and 
rectangular grooves, $l_g(x)$ (full lines), for the postfilling scaling regime. Black, red, and blue colors
correspond to $(R = 50, P =110)$, $(R = 120, P =250)$, and $(R = 245, P =500)$,
respectively. The dashed and full interfacial profiles  are calculated for two different values of the undersaturation
which both satisfy  $C_{p,g}\Delta\mu/\varepsilon_f(R/\sigma)^{1.035}=1.18$ with $C_p=1$ and $C_g=2$, respectively. In the inset the axes are
rescaled by $R^{-1}$ leading to  complete data collapse. 
 $D=500$ and $\sigma$ is the length unit; $A/\varepsilon_f \approx  0.330$ and $\sigma_{lg}\sigma^2/\varepsilon_f\approx 0.478$.}
\label{profiles-geom-covar}
\end{figure}

 In the postfilling scaling regime, in addition to the midpoint interfacial heights, those values 
of the profiles $l_{p,g}({\bf x})$, which are taken off center at the
same distance $r_{p,g}/R\lesssim 1$ from the symmetry axis of a pit or from the midplane of a groove,
also satisfy generalized scaling relations  similar to Eqs.~(\ref{eq:main_1a}) and (\ref{eq:main_1b}):
\begin{eqnarray}
l_{p}({\bf x },\Delta\mu,R,P,D)=R\Lambda_{p}\biggl (\frac{r_p}{R},\frac{\Delta\mu}{\varepsilon_f} \biggr(\frac{R}{\sigma}\biggl)^{1+\delta} \biggr ) \\
l_{g}(x,\Delta\mu,R,P,D)=R\Lambda_{g}\biggl (\frac{r_g}{R},\frac{\Delta\mu}{\varepsilon_f} \biggr(\frac{R}{\sigma}\biggl)^{1+\delta} \biggr ),
\label{eq:main_2a}
\end{eqnarray}
where $r_p = \sqrt{(x-x_{0})^2+(y-y_{0})^2}$ is the radial distance from the symmetry axis of a pit at $(x,y)=(x_0,y_0)$;
$r_g=|x-x_{0}|$ is the distance from the midplane of a groove at $x=x_0$. The scaling functions $\Lambda_{p}$ and $\Lambda_{g}$ 
obey the covariance relation
\begin{equation}
\Lambda_p(u,v)=\Lambda_g(u,v/2), \hspace*{0.4cm}u\lesssim 1.
\label{eq:main_2b}
\end{equation}
The interfacial profiles for cylindrical pits, $l_p$, and rectangular grooves, $l_g$,  are shown in the Fig.~\ref{profiles-geom-covar}
for several  $R$. The profiles are calculate for two different undersaturations $\Delta\mu$ within the postfilling regime which both satisfy 
$C_{p,g}\Delta\mu/\varepsilon_f(R/\sigma)^{1.035}=1.18$ with $C_p=1$ and $C_g=2$, respectively. Rescaling the variables according to
Eq.~(\ref{eq:main_2a}) leads to data collapse. Equation~(\ref{eq:main_2b})  means that, for fixed $R$ and certain values of undersaturation, the interfacial profile
within a pit region can be considered as a surface of revolution generated by the function $l_g(r_g)$
which characterizes the interfacial profile within a groove region. This  also holds within the
filling scaling regime.  

 \begin{figure}[]
\begin{center}
\includegraphics[width=0.8\linewidth]{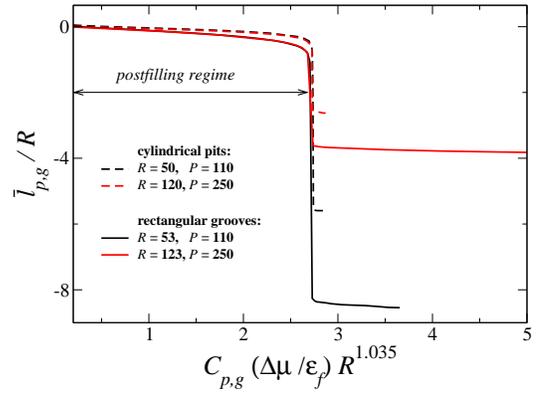}
\end{center}
\caption{Rescaled laterally averaged interfacial profile for  cylindrical pits,  $\bar l_{p}$ (dashed lines), and for
rectangular grooves, $\bar l_{g}$ (full lines), as a function of 
undersaturation  $\Delta\mu$. The rescaling of the variables  according to 
Eqs.~(\ref{eq:main_1a}) and (\ref{eq:main_1b}) leads to two different scaling 
curves describing  the cylindrical pits and the rectangular grooves, respectively. 
Thus the covariance relation given by Eq.~(\ref{eq:main_1b}) is not fulfilled.
Compare Fig.~\ref{l0_rect}(b) which shows that the postfilling scaling functions for the 
midpoint interfacial height are the same for these two cavity types. $C_p=1$ and $C_g=2$, $D=500$, and
$\sigma$ is the length unit; $A/\varepsilon_f \approx  0.330$ and $\sigma_{lg}\sigma^2/\varepsilon_f\approx 0.478$. }
\label{l_aver_rect}
\end{figure}

\begin{figure}[]
\begin{center}
\includegraphics[width=0.8\linewidth]{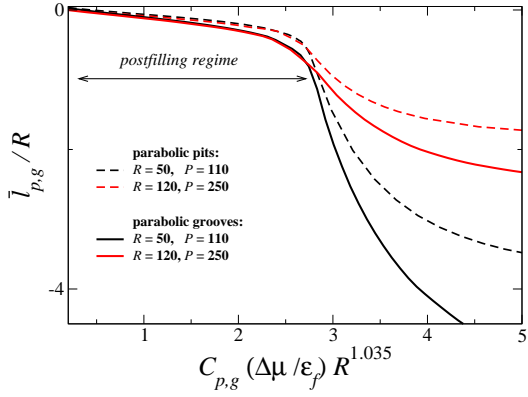}
\end{center}
\caption{Rescaled laterally averaged interfacial profile for parabolic pits, $\bar l_{p}$ (dashed lines),
and grooves, $\bar l_{g}$ (full lines), as a function of 
undersaturation  $\Delta\mu$. The rescaling of the variables  according to 
Eqs.~(\ref{eq:main_1a}) and (\ref{eq:main_1b}) leads to two different scaling 
curves describing  the cylindrical pits and the rectangular grooves, respectively. 
Compare Fig.~\ref{l0_parab}(b) which shows that the postfilling scaling functions
for the midpoint interfacial height are the same for these two cavity types. 
 $C_p=1$ and $C_g=2$, $D=500$, and
$\sigma$ is the length unit; $A/\varepsilon_f \approx  0.330$ and $\sigma_{lg}\sigma^2/\varepsilon_f\approx 0.478$.}
\label{l_aver_parab}
\end{figure}

We also consider the  lateral average  $\bar l_{p,g} = \int d^2x\hspace*{0.2cm}l_{p,g}({\bf x})/\bar{{\cal A}}$ 
of the interfacial profiles $l_{p,g}({\bf x})$  with the integral running over the lateral surface area $\bar{{\cal A}}$ of 
a  unit cell of the surface pattern.  
This quantity is related to the excess adsorption $\Gamma = \int d^3r (\rho({\bf r})-\rho_b)/\bar{{\cal A}}$ 
per area of the cell, with the integral running over the region accessible to the fluid above a unit cell.
Using the sharp-kink approximation for the density (see Eq.~(\ref{kink})) $\Gamma$ can be expressed as 
\begin{eqnarray}
\Gamma_{p,g} &=& \frac{1}{\bar{{\cal A}}}\Delta\rho \int\limits_{\bar{{\cal A}}} d^2x \Bigl (l({\bf x}) - s(\bf{x})\Bigr ) \nonumber \\
&=& \Delta\rho\Bigl (\bar l_{p,g} + \frac{V_{p,g}}{\bar{{\cal A}}}\Bigr ),
\label{adsorption_kink_approx}
\end{eqnarray}
where $V_{p,g}$ is the volume of a pit or groove, respectively.
For rectangular  grooves and cylindrical pits Fig.~\ref{l_aver_rect} shows the corresponding results
for $\bar l_{p,g}(\Delta\mu)$ with the chemical potential
and $\bar{l}$ rescaled according to Eqs.~(\ref{eq:main_1a}) and (\ref{eq:main_1b}). The laterally averaged interfacial 
profile in the postfilling scaling regime also obeys the scaling described by Eq.~(\ref{eq:main_1a}), but the scaling functions 
for the rectangular grooves and the cylindrical pits do {\em not} satisfy the covariance relation  given by Eq.~(\ref{eq:main_1b}).
The same result holds for the case of parabolic pits and grooves, which is shown in Fig.~\ref{l_aver_parab}. Moreover, the
scaling functions which are shown in Fig.~\ref{l_aver_rect}, describe also the laterally averaged
interfacial profiles for parabolic  pits and grooves, respectively.
Geometrical considerations show that the lateral average of a troughlike surface
generated by a function of a single variable is smaller than the  lateral average of the corresponding 
rotationally symmetric  surface generated by revolving the same function around the symmetry axis. 
As shown above, both surfaces obey the covariance scaling properties. Therefore this
inequality  of the lateral averages  implies that this covariance does not hold for the lateral averages.

\subsection{Small $\Delta\mu:$ effective planar and planar scaling regimes}

\begin{figure}[]
\begin{center}
\includegraphics[width=0.7\linewidth]{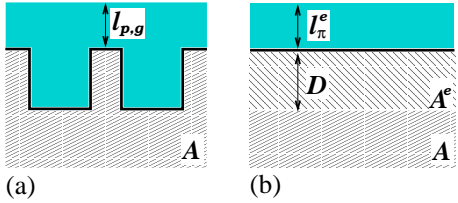}
\end{center}
\caption{(a) Schematically drawn liquid wetting film morphology on substrates sculptured by cylindrical
pits or rectangular grooves within the  effective
planar scaling regime. 
(b) Wetting  film of thickness $l_{\pi}^e$ on a flat layered solid. The top layer
has a thickness $D$ with an effective Hamaker constant $A^e$ whereas the semi-infinite bottom 
part of the substrate is characterized by the Hamaker constant $A$.}
\label{layered_solid}
\end{figure}

\begin{figure}[]
 \begin{center}
\includegraphics[width=0.8\linewidth]{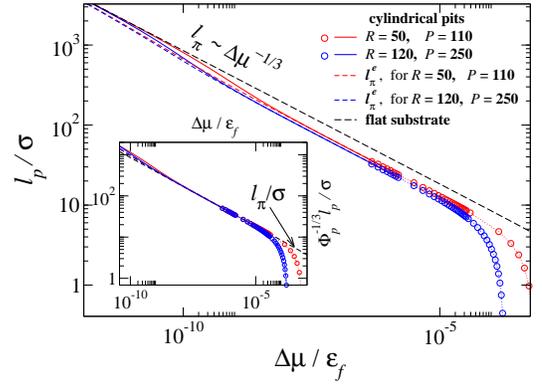}
\end{center}
\caption{Wetting film thickness $l_p$ above cylindrical pits for small $\Delta\mu$.
The solid lines are obtained by minimizing Eq.~(\ref{eff_hamilton_local}) in the subspace of laterally constant 
interfacial heights  $l({\bf x })=l_p$. 
The circles represent the mid-point interfacial heights $l_p^{(0)}$ and
are the results of the full numerical minimization of the Hamiltonian in
Eq.~(\ref{eff_hamilton_local}). The circles are plotted in that $\Delta\mu$ range for which the deviations
from the full and dashed lines become detectable.
The dashed lines correspond to the film thicknesses $l_{\pi}^e(\Delta\mu)$ on  layered flat ersatz substrates
(see Fig.~\ref{layered_solid}), with a top layer of height $D$ and
an effective Hamaker constant $A^e=A\Phi_p\equiv A(1-\pi(R/P)^2)$, where $A$ is the  Hamaker constant of the solid without pits.
In the inset the vertical axis is rescaled according to the scaling
relation given in Eq.~(\ref{eq:gc_planar_H}) leading to data collapse and -- after rescaling -- agreement with $l_{\pi}$ 
within an intermediate regime,  $\Delta\mu_{\pi}<\Delta\mu<\Delta\mu_{\pi}^e$,
the width of which increases as
 $\Delta\mu_{\pi}^e-const\times D^{-3}$ for $D\rightarrow\infty$.
The  pit depth is $D=500\sigma$; $A/\varepsilon_f \approx  0.330$ and $\sigma_{lg}\sigma^2/\varepsilon_f\approx 0.478$.}
\label{lp_cyl}
\end{figure}

\begin{figure}[]
\begin{center}
\includegraphics[width=0.8\linewidth]{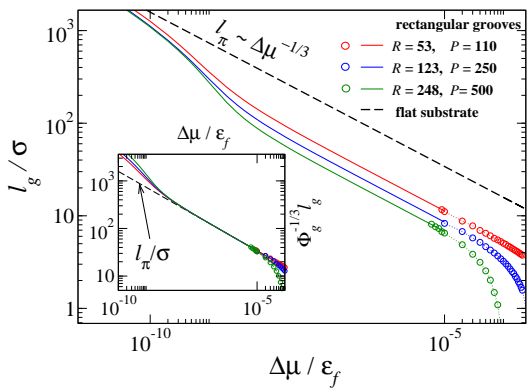}
\end{center}
\caption{Wetting film thickness $l_g$ above rectangular groove patterns for small $\Delta\mu$.
The solid lines are obtained by minimizing Eq.~(\ref{eff_hamilton_local}) in the subspace of laterally constant 
interfacial heights  $l({\bf x })=l_g$. The circles represent the mid-point interfacial heights $l_g^{(0)}$ and
are the results of the full numerical minimization of the Hamiltonian in
Eq.~(\ref{eff_hamilton_local}). The circles are plotted in that $\Delta\mu$ range for which the deviations
from the full and dashed lines become detectable.
On the present scales the data for $l_g$ are indistinguishable from the lines $l_{\pi}^e(\Delta\mu)$ obtained 
for layered flat ersatz substrates with a top layer of height $D$ and
an effective Hamaker constant $A^e=A\Phi_g\equiv A(1-2R/P)$, where $A$ is the  Hamaker constant of the solid without grooves.
In the inset the vertical axis is rescaled according to the scaling
relation given in Eq.~(\ref{eq:gc_planar_H}) leading to data collapse and -- after rescaling -- agreement with $l_{\pi}$
 within an intermediate regime, $\Delta\mu_{\pi}<\Delta\mu<\Delta\mu_{\pi}^e$,
the width of which increases as
 $\Delta\mu_{\pi}^e-const\times D^{-3}$ for $D\rightarrow\infty$.
The  groove depth is $D=500\sigma$; $A/\varepsilon_f \approx  0.330$ and $\sigma_{lg}\sigma^2/\varepsilon_f\approx 0.478$.}
\label{lg_grooves}
\end{figure}

For undersaturations below a certain value $\Delta\mu_{\pi}^e$, i.e., closer to gas-liquid coexistence,
 the liquid-gas interface becomes flat (Fig.~\ref{layered_solid}), its height denoted as $l_{p,g}$ for the pit and groove 
patterns, respectively.
For these flat wetting films the functional in Eq.~(\ref{eff_hamilton_local}) reduces to a function of
a single variable $l_{p,g}$ the minimum of which determines the equilibrium thicknesses of the wetting films in this 
regime. The corresponding results  $l_{p,g}(\Delta\mu)$ for  cylindrical pits and rectangular grooves
are presented in Fig.~\ref{lp_cyl} and Fig.~\ref{lg_grooves}, respectively. 
For these two substrate geometries there is an interval of undersaturations 
$(\Delta\mu_{\pi},\Delta\mu_{\pi}^e)$, within which both $l_p$ and $l_g$ reveal the same power law growth 
$l_{p,g}\sim\Delta\mu^{-1/3}$
as for a flat surface, but with different amplitudes reflecting
different, geometry dependent, {\em e}ffective Hamaker constants $A^e$. 
We call this range of undersaturation the effective planar scaling regime.
 Within  this regime the  substrates sculptured by cavities with vertical walls are equivalent to layered and flat ersatz solids 
(see Fig.~\ref{layered_solid}(b))
with the effective interface potential \cite{robbins:91}
\begin{equation}
W(l\gg\sigma)\approx\frac{\pi}{12}\biggl (\frac{A}{(l+D)^2} + \frac{A^e}{l^2} - \frac{A^e}{(l+D)^2}\biggr ).
\label{eq:layered_solid}
\end{equation}
Here the first term corresponds to a flat semi-infinite, $z\in(-\infty,-D)$,  solid with Hamaker constant $A$  and the
second and the third are the corrections due to the finite slab, $z\in(-D,0)$, with effective Hamaker constant $A^e$.
For $\sigma\ll l\lesssim D$, it is a good approximation to ignore the bottom part of the substrate 
(i.e., the first and the third terms in $W(l\gg\sigma)$) so that the
 amplitude is determined by the effective Hamaker constant $A^{e}$.

With $\Lambda_s$ we denote that region of the sculptured part of the substrate ($z\in[-D,0]$), which is
occupied by the solid, and with $\Lambda_0$ the entire slab $z\in[-D,0]$.
The equivalence of the patterned and the flat layered substrate 
actually  means $W_{\Lambda_s}({\bf x},l)=W^{e}_{\Lambda_0}({\bf x},l)$. With Eq.~(\ref{interface_potential-2}) 
one obtains $A^{e}=A I_{\Lambda_s}/ I_{\Lambda_0}$. 
For $l\gg\sigma$ the integrals $I_{\Lambda_i}$ ($i=s,0$) can be approximated as 
\begin{equation}
I_{\Lambda_i} \approx \int\limits_{l}^{\infty}dz\int\limits_{\Lambda_i}\frac{d^3r^{\prime}}{(z-z^{\prime})^6}
\biggl (1-3\frac{\sigma^2+||{\bf x}-{\bf x}^{\prime}||^2}{(z-z^{\prime})^2 } \biggr). 
\label{eq:int_ratio}
 \end{equation}

\noindent Actually, $I_{\Lambda_s}$  is a functions of ${\bf x}$, which varies across a unit cell
and only its leading term $\sim l^{-2}$ 
is laterally constant. In Eq.~(\ref{eq:int_ratio}) taking into account  
only the leading term as function of $l$  
one finds  $A^{e}\approx A S_s/S_0$, where $S_s/S_0$ is the  fraction of area occupied by the solid in the top layer
of the substrates patterned by cavities with vertical walls.  Thus, one has $A^{e}_g\approx A \Phi_g$ for the rectangular grooves 
and  $A^{e}_p\approx A \Phi_p$
for  cylindrical pits (see Eq.~(\ref{eq:gc_planar_H})). 
The thicknesses  $l_{\pi}^{e}(\Delta\mu)$ of the wetting films   
on such layered ersatz substrates, which effectively correspond to  arrays of grooves, are 
 almost indistinguishable from the corresponding actual ones $l_{g}$. 
For lattices of cylindrical pits in Fig.~\ref{lp_cyl} we present  $l_{\pi}^{e}(\Delta\mu)$ (dashed lines)
 on layered substrates,  and find agreement with the actual thicknesses $l_p(\Delta\mu)$ within the effective planar
scaling regime. Therefore we can conclude that for sufficiently  thick wetting films $l_{p,g}$ obey
the scaling relation given in Eq.~(\ref{eq:gc_planar_H}).  Indeed, after rescaling the film thicknesses 
by the geometry-dependent factors $(\Phi_{p,g})^{-1/3}$, the curves for $l_{p,g}$ collapse and, within the numerical precision, 
they coincide with the  curve for $l_{\pi}$. This is shown in the insets  of  Figs.~\ref{lp_cyl} and \ref{lg_grooves}. 
In the case of parabolic cavities we have found numerically that the corresponding  curves 
for $l_{p,g}$ do not reveal this effective planar scaling regime, but
gradually cross over to the  planar one, denoted as $l_\pi$ \cite{tasinkevych:06:0}. 
We emphasize that the effective planar regime exists {\it only} for patterns with {\em vertical} walls. Thus, we also
expect this regime to exist for lattices of cylindrical nanoposts,  but not for, e.g., sawlike substrate surfaces
or periodic arrays of trapezoidal cavities.

As  $l_{p,g}$  reach the value $\sim D$, i.e., at $\Delta\mu=\Delta\mu_{\pi}\sim D^{-3}$,  
a crossover to the  planar scaling regime takes place. In the planar scaling 
regime $\Delta\mu\lesssim\Delta\mu_{\pi}$ the geometrical patterns have become irrelevant. 
Thus the range of  applicability $\Delta\mu_{\pi}^e-\Delta\mu_{\pi}$ of Eq.~\ref{eq:gc_planar_H} increases with the cavity depth $D$ as
$\Delta\mu_{\pi}^e-const\times D^{-3}$, whereas
for $D\rightarrow 0$ \hspace*{0.2cm} $\Delta\mu_{\pi}$  merges  
with  $\Delta\mu_{\pi}^{e}$, eliminating the effective planar scaling regime.

\section{Comparison with experimental data}
\label{theory_vs_experiment}
\begin{figure}[]
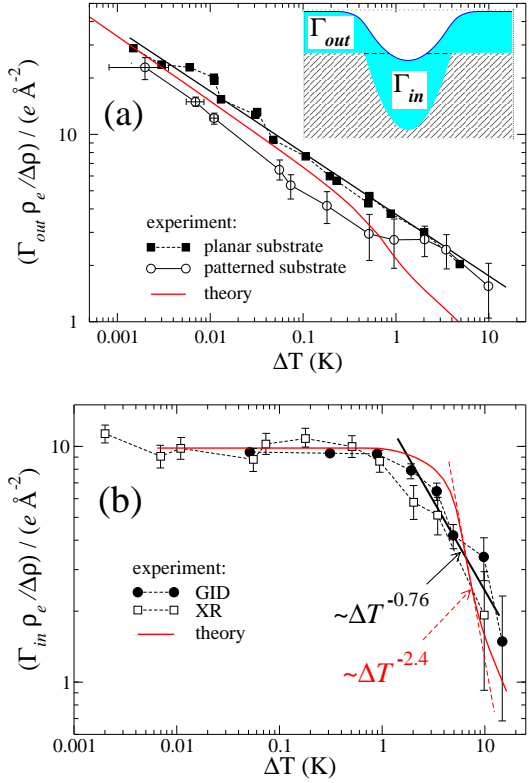

\begin{center}
\includegraphics[width=0.8\linewidth]{fig13a.eps}
\end{center}
\begin{center}
\includegraphics[width=0.8\linewidth]{fig13b.eps}
\end{center}
\caption{  Excess adsorptions (a) $\Gamma_{out}(\Delta T)$ above the top of the substrate and 
(b) $\Gamma_{in}(\Delta T)$ within the parabolic pits.  
Symbols and black lines denote experimental data from Ref.~\cite{gang-exper_prl:05}.
Red lines are the corresponding theoretical results (see main text).
$\rho_e=0.26 e/$\AA$^3$ is the electron density of the bulk MCH.
$\Delta T$ is proportional to $\Delta\mu$.}
\label{experiment}
\end{figure}

In this section we compare our results with the corresponding experimental data of Ref.~\cite{gang-exper_prl:05}, where
the adsorption of methylcyclohexane (MCH) on a nanopatterned silicon substrate has been studied using X-ray reflectivity (XR) 
and grazing incident diffraction (GID) techniques.
The substrate was patterned by a hexagonal lattice of parabolic pits with radius $R\approx 123$\AA \hspace*{0.1cm}, 
depth $D\approx 200$\AA \hspace*{0.1cm}, and  lattice constant $P\approx 394$\AA. Accordingly, in the following the theoretical
data have been obtained for a hexagonal pattern, too. The data in  Ref.~\cite{gang-exper_prl:05} have been interpreted in terms of the excess adsorption $\Gamma=\Gamma_{in}+\Gamma_{out}$
as the sum of the excess adsorption $\Gamma_{in}$  in the pit and  the excess adsorption $\Gamma_{out}$ above the pit opening
(see inset in Fig.~\ref{experiment}(a)).
These two quantities are defined as
\begin{equation}
\Gamma_{in} = \frac{1}{\bar{{\cal A}}}\int\limits_{\bar{{\cal A}}} d^2x \int\limits_{s({\bf x})}^0dz\Bigl( \rho({\bf r}) - \rho_g \Bigr)
\label{gama_p-0}
\end{equation}
and
\begin{equation}
\Gamma_{out} = \frac{1}{\bar{{\cal A}}}\int\limits_{\bar{{\cal A}}} d^2x \int\limits_{0}^{\infty}dz\Bigl( \rho({\bf r}) - \rho_g \Bigr).
\label{gama_t-0}
\end{equation}
Using the sharp-kink approximation for the density profile (see Eq.~(\ref{kink})) Eqs.~(\ref{gama_p-0}) and (\ref{gama_t-0}) take the form
\begin{equation}
\Gamma_{in} = \frac{\Delta\rho}{\bar{{\cal A}}}\int\limits_{\bar{{\cal A}}}d^2x \Bigl( \Theta( -l({\bf x}) ) l({\bf x})-s({\bf x}) \Bigr),
\label{gama_p}
\end{equation}
and
\begin{equation}
\Gamma_{out} = \frac{\Delta\rho}{\bar{{\cal A}}}\int\limits_{\bar{{\cal A}}} d^2x \hspace*{0.2cm}\Theta( l({\bf x})) l({\bf x}).
\label{gama_t}
\end{equation}
In these experiments the thermodynamic deviation from liquid-vapor coexistence of MCH
has been tuned via controlling a temperature difference $\Delta T$ between the substrate and the reservoir of bulk liquid.
 $\Delta T$ is related to the undersaturation
$\Delta\mu$ according to  \cite{gang-exper_prl:05} $\Delta T = \Delta\mu T / H$, where
$T=305 K$ is the reservoir temperature, and $H = 5.8\times10^{-20}J$ is  the latent heat of vaporization per
MCH molecule \cite{gang-exper_prl:05}. We have modeled MCH by a Lennard-Jones interaction potential with the parameters  
$\sigma=5.511${\AA} and $\varepsilon_f/k_B=446 K$ \cite{goldman}, which leads to 
 $\Delta T\approx 32.37 \frac{\Delta\mu}{\varepsilon_f} K$.
By fitting the experimental reference data for the wetting film thickness on the corresponding planar substrate
 \cite{gang-exper_prl:05}  to the expression $l_{\pi}(\Delta\mu\rightarrow 0)$ given just below 
Eq.~(\ref{interface_potential_planar}) 
we are able to determine the ratio $A/(\varepsilon_f \Delta\rho^*)\approx 1.006$ involving the Hamaker constant $A$. 
The last material parameter required to facilitate  a
quantitative comparison between our theory
and the experimental data in Ref.~\cite{gang-exper_prl:05} is  the liquid-vapor surface tension of MCH, for which
we use  $\sigma_{lg}=22.72\frac{dyn}{cm}$ for  $T=30^{\circ}C$   \cite{jasper}.
 Using the values of $\sigma$ and
$\varepsilon_f$ given above for MCH, we find  $ \sigma_{lg} \sigma^2/\varepsilon_f \approx 1.149$. According to 
Eq.~(\ref{surface_tension}) this value of the surface tension corresponds to $\Delta\rho^*\approx 0.668$ or $T^*\approx0.88$
as follows from the bulk phase digram shown in Fig.~\ref{bulk_diagram}.  It is satisfactory to see that this value of the
reduced  temperature $T^*/T^*_c = T/T_c \approx 0.88/1.41 \approx 0.62$ turns out to be reasonably close to the actual experimental
ratio $T/T_c=305/572\approx 0.53$.
With this there are no free parameters left in the model.

$\Gamma_{out}$ and $\Gamma_{in}$ as functions of $\Delta T$ are shown in Fig.~\ref{experiment}(a) and  Fig.~\ref{experiment}(b), respectively. 
The $X$-ray reflectivity technique, which has been used in Ref.~\cite{gang-exper_prl:05}, provides
access to the laterally averaged  electron-density profile $\rho(z)$; consequently the adsorption is measured  in  units of
the areal electric charge density $e$\AA$^{-2}$. In order to present the theoretical adsorption data
in these units, we use for the electron-density of the bulk liquid MCH the value $\rho_e = 0.26 e$\AA$^{-3}$ and neglect
the density of the vapor.
For temperature differences in the range $5K \lesssim\Delta T\lesssim 8K$ we identify  the filling regime, i.e.,
within this range the theoretical data $\Gamma_{in}(\Delta T)$ resemble a power-law growth $\Gamma_{in}\propto\Delta T^{-\beta_p}$ with 
an {\it effective} exponent 
$\beta_p\approx 2.4$. This effective exponent $\beta_p$  turns out to increase with the {\it p}it size and for  
$R/\sigma=245$ and $D/\sigma=500$ we find  $\beta_p \simeq 3.4$.
There is no obvious connection between the effective exponent $\beta_p$ and the exponent $\gamma(R,D)$, which is introduced in  Subsec.~\ref{results-large-delta-mu}
for the midpoint interfacial height.  In Ref.~\cite{gang-exper_prl:05} the value of effective exponent $\beta_p \simeq 3.4$
has been roughly estimated by  using $\gamma=2$ (which is valid for infinitely deep parabolic wedges or pits) and the expression
 $V(l_p^{(0)}) =  \frac{\pi}{2} ( R l_p^{(0)})^2 /D + \frac{\pi }{3}(R^2 l_p^{(0)}/D)^{3/2}$ for the volume
enclosed between the surface of parabolic pit and a spherical meniscus \cite{gang-exper_prl:05,parry_nature:00} which
approximates the shape of the interfacial profile.  With this, for the 
experimentally relevant values of $l_p^{(0)}$ the dependence of $V$ on $\Delta\mu$  can be 
approximated as $V(l_p^{(0)}\sim \Delta\mu^{-2})  \sim \Delta\mu^{-3.4}$ \cite{gang-exper_prl:05}.

At $\Delta T\approx 8K$ one can identify a small change in  slope of the  $\Gamma_{in}(\Delta T)$ curve. This crossover
corresponds to the disappearance of the interfacial meniscus, such that for larger values 
of $\Delta T$ the interface follows the substrate shape.  
The authors of  Ref.~\cite{gang-exper_prl:05} assigned the value $\beta_p\approx 0.76$ to their data. We explain this disagreement with 
the value $\beta_p\approx 2.4$ reported here (see Fig.~\ref{experiment}(b)) by a mislocation of the filling regime for the experimental
data of  $\Gamma_{in}(\Delta T)$ induced by the large error bars for the data points at large $\Delta T$.
There also seems to be a difference between the behavior 
of the theoretical and experimental results for $\Gamma_{out}$ at large $\Delta T$ (see Fig.~\ref{experiment}(a)).   
However, for such large undersaturations the wetting film is only ca. $10${\AA} 
 thick causing rather large experimental error bars
 (actually larger \cite{private_comm} than those presented 
 in  Fig.~\ref{experiment}(a) which are taken from Ref.~\cite{gang-exper_prl:05}) 
for  extracting values for the adsorption $\Gamma_{out}$ from the scattering data.  
On the other hand, for such thin wetting films the reliability of our theoretical
description is expected to deteriorate, too. Due to packing effects the use 
of an effective interface Hamiltonian should be replaced by a full-fledged density 
functional approach.

\begin{figure}[]
\begin{center}
\includegraphics[width=0.8\linewidth]{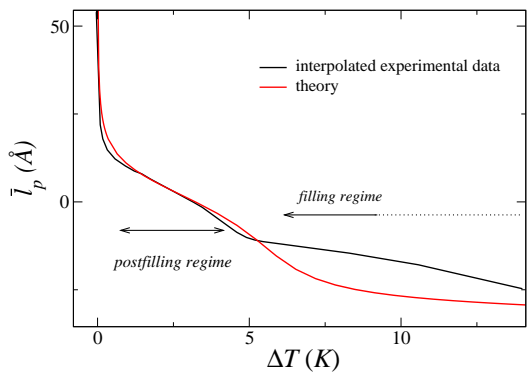}
\end{center}
\caption{Laterally averaged interfacial profile $\bar l_p(\Delta T)$ for parabolic pits, 
obtained according to Eq.~\ref{adsorption_kink_approx}. The black line  
interpolates the experimental data from Ref.~\cite{gang-exper_prl:05} (without error bars), and
the red line denotes corresponding theoretical results. $\Delta T$ is proportional to $\Delta\mu$
(see the main text after Eq.~(\ref{gama_t})).}
\label{l_aver_parab_experiment}
\end{figure}

In Fig.~\ref{l_aver_parab_experiment} the laterally averaged interfacial profile $\bar l_p(\Delta T)$ is shown.
In order to obtain the experimental curve $\bar l_p$,  first we have interpolated the experimental data for $\Gamma_{in}$ and
$\Gamma_{out}$, second we have formed $\Gamma=\Gamma_{in} + \Gamma_{out}$, and by using Eq.~(\ref{adsorption_kink_approx})
we have determined $\bar l_p(\Delta T)$. This experimental curve is compared with the corresponding theoretical one.
The theoretical curve for $\bar l_p$  reveals the filling and postfilling scaling regimes, as well as the extended crossover to the planar
scaling regime. Due to the relatively small depth $D/\sigma\approx 36$ of the pits the crossover region
between the first two regimes is rather extended.
In the postfilling scaling regime, i.e., for $0.7K \lesssim\Delta T\lesssim 4K$,
we obtain very good agreement between the experimental and theoretical data, while there are substantial
deviations in the filling regime. 
With respect to the latter one we suggest that for deeper pits the experimental identification of the filling regime would be easier and the error bars
smaller. First under this conditions  the filling regime occurs over a wider range
of undersaturations; second, the midpoint interfacial height attains larger values; third, the crossover to the postfilling regime becomes more pronounced.  

\section{Summary and outlook}
\label{summary}
Within an effective interface Hamiltonian approach based on density functional theory (Sect.~\ref{model})
we have studied complete wetting of four classes of solid substrates topographically patterned on the nanoscale: 
quadratic (and hexagonal) lattices of cylindrical or parabolic pits and periodic arrays of rectangular or parabolic grooves (Fig.~\ref{geometry}).
The long-ranged fluid-fluid interaction (giving rise to the bulk phase diagram shown in Fig.~\ref{bulk_diagram}) and the fluid-substrate potential
 are modelled by  isotropic pair
potentials decaying with distance $\sim 1/r^6$.
By analyzing the full range of undersaturations $\Delta\mu$ relative to  liquid-vapor coexistence in the bulk we have
identified four different regimes for the increase of the wetting film thickness
(Subsecs.~\ref{results} A and \ref{results} B), and the existence 
of so-called covariances relating this increase for pits and grooves (Eq.~(\ref{eq:main_1b})).

The filling of a cavity (Fig.~\ref{profiles}) by the liquid occurs independently 
of the  presence of other ones. This suggests 
that, e.g.,  in  the filling regime the power law increase of the excess coverage in  parabolic cavities
is not only accessible to X-ray scattering (which requires periodic arrays) but
also to AFM techniques applied to  single  cavities (see, e.g., Ref.~\cite{fukuzawa:04}).
For $D/R\gtrsim 0.5$ (see Fig.~\ref{geometry}) there exists a range of undersaturations, which we call the postfilling scaling regime,
in which the local interfacial characteristics, obtained for all four patterns, can be expressed in terms of a {\em single} 
scaling function  (Eqs.~(\ref{eq:main_1a}), (\ref{eq:main_1b}), (\ref{eq:main_2a}), and  (\ref{eq:main_2b}) and 
Figs.~\ref{l0_rect}, \ref{l0_parab}, and \ref{profiles-geom-covar}). 
However, the characterization of the laterally  averaged interfacial profile requires {\em two} different scaling 
functions,  one for pits and one for grooves (Figs.~\ref{l_aver_rect} and \ref{l_aver_parab}).  All these scaling functions are
independent of the cavity depth (Fig.~\ref{l0_cyl_eff_depth}).

For small undersaturations, the increase of the wetting film thickness is determined  by the entire array of cavities.
In the case of vertical cavity  walls, i.e., for rectangular grooves and cylindrical pits, the interfacial thickness increases as on a planar substrate,
but with an effective,  geometry-dependent, Hamaker constant (Eq.~\ref{eq:gc_planar_H} and Figs.~\ref{layered_solid}-\ref{lg_grooves}). 
For very small undersaturations the patterns on the substrate become irrelevant, and the 
wetting film thickness increases as if the substrate would be flat. 
The crossover to this latter so-called planar scaling regime occurs only for  long-ranged dispersion forces. 
For  short-ranged interactions, instead, the  increase of the wetting film thickness would remain  determined by 
the areal fraction of the solid at the substrate surface $z=0$ for {\em all} film thicknesses \cite{harnau:04}. 

We have been able to report  quantitative agreement between our theory and the $X-$ray scattering data of Ref.~\cite{gang-exper_prl:05}
(Figs.~\ref{experiment} and \ref{l_aver_parab_experiment}). For the range of undersaturations
corresponding to the postfilling scaling regime, the agreement is satisfactory  (Fig.~\ref{l_aver_parab_experiment}).
We have been able to clarify the interpretation of the adsorption  parabolic pits in terms of an effective filling exponent $\beta_p$ 
(Fig.~\ref{experiment}(b)). 
We suggest, that for deeper pits the experimental identification of the filling regime would be easier. 
In order to test our scaling predictions (Eqs.~(\ref{eq:main_1a}) and (\ref{eq:main_1b}))
 experiments  for the following surface structures are still missing:
(i)  patterns with  cavities of different sizes have to be considered;
(ii) resolving the full morphology of the wetting films, or at least determining the midpoint interfacial
heights, are highly desirable.

A natural extension of the work presented here would be to study the temperature dependence
of the filling regime and of the scaling function for the postfilling scaling regime. Another 
perspective is to consider substrates patterned by, e.g., cylindrical nanoposts.

\acknowledgments 
M.T. gratefully acknowledges helpful discussions with M. N. Popescu.


\begin{thebibliography}{99}






\bibitem{gang-exper_prl:05}
O. Gang, K. J. Alvine, M. Fukuto, P. S. Pershan, C. T. Black, and B. M. Ocko,
Phys. Rev. Lett. {\bf 95}, 217801 (2005).

\bibitem{bruschi-exper_prl:02}
L. Bruschi, A. Carlin, and G. Mistura,
Phys. Rev. Lett. {\bf 89}, 166101 (2002).

\bibitem{Bruschi:06:1}
 L. Bruschi, G. Fois, G. Mistura, M. Tormen, V. Garbin, E. di Fabrizio, A. Gerardino, and M. Natali,
J. Chem. Phys. {\bf 125 },  144709 (2006).


\bibitem{reimer:99}
K. Rejmer, S. Dietrich, and M. Napi\'{o}rkowski,
Phys. Rev. E {\bf 60} 4027 (1999).

\bibitem{parry_nature:00}
C. Rasc\'{o}n and  A. O. Parry, Nature {\bf 407}, 986 (2000);

\bibitem{parry_jcp:00}
C. Rasc\'{o}n and  A. O. Parry, J. Chem. Phys. {\bf 112}, 5157 (2000).





\bibitem{parry_prl:05} 
C. Rasc\'{o}n  and A. O. Parry, 
Phys. Rev. Lett. {\bf 94}, 096103 (2005).

\bibitem{tasinkevych:06:0}
M. Tasinkevych and S. Dietrich, Phys. Rev. Lett. {\bf 97}, 106102 (2006). 


\bibitem{dietrich_book}
S. Dietrich, in {\it Phase Transitions and Critical Phenomena,}
 edited by C. Domb and J. L. Lebowitz (Academic, London, 1988), Vol. 12, p. 1.

\bibitem{dietrich:91}
S. Dietrich  and M. Napi\'{o}rkowski,
Phys. Rev. A {\bf 43} 1861 (1991).

\bibitem{darbellay_groove:92}
G. A. Darbellay and J. M. Yeomans,
J.  Phys. A: Math. Gen. {\bf 25}, 4275 (1992).

\bibitem{robbins:91}
M. O. Robbins, D. Andelman, and J. F. Joanny,
Phys. Rev. A {\bf 43}, 4344 (1991).

\bibitem{napiorkowski_wedge:92}
M. Napi\'{o}rkowski, W. Koch, and S. Dietrich,
Phys. Rev. A {\bf 45}, 5760 (1992).

\bibitem{koch:95}
W. Koch, S. Dietrich, and  M. Napi\'{o}rkowski, Phys. Rev. E {\bf 51}, 3300 (1995).

\bibitem{bauer_dietrich}
C. Bauer and S. Dietrich,
Phys. Rev. E {\bf 60}, 6919 (1999); {\bf 61}, 1664 (2000); 
C. Bauer, S. Dietrich, and A. O. Parry, Europhys. Lett. {\bf 47}, 474 (1999).





\bibitem{evans:79}
R. Evans, Adv. Phys. {\bf 28}, 143 (1979). 

\bibitem{WCA}
J. D. Weeks, D. Chandler, and H. C. Andersen,
J. Chem. Phys. {\bf 54}, 5237 (1971).



\bibitem{napiorkowski:91:0}
 S. Dietrich and M. Napi\'{o}rkowski, Physica A {\bf 177}, 437 (1991);
M. Napi\'{o}rkowski and S. Dietrich, Z. Phys. B {\bf 89}, 263 (1992);
Phys. Rev. E {\bf 47}, 1836 (1993).


\bibitem{carnahan:69}
N. F. Carnahan and K. E. Starling,
J. Chem. Phys. {\bf 51}, 635 (1969).

\bibitem{mecke:99}
K. R. Mecke and S. Dietrich, Phys. Rev. E {\bf 59}, 6766 (1999).

\bibitem{bauer:99}
C. Bauer and S. Dietrich, Eur. Phys. J. B {\bf 10}, 767 (1999).




\bibitem{goldman} 
S. Goldman, J. Phys. Chem. {\bf 80}, 1697 (1976). 

\bibitem{jasper} J. J. Jasper, J. Phys. Chem. Ref. Data, {\bf 1}, 841 (1972).
\bibitem{private_comm} O. Gang, P. S. Pershan, and B. M. Ocko, private communication;
see also O. Gang, K. J. Alvine, M. Fukuto, P. S. Pershan, C. T. Black, and B. M. Ocko,
Phys. Rev. Lett. {\bf 97}, 039902 (2006).
\bibitem{fukuzawa:04} K.~Fukuzawa, J.~Kawamura, T.~Deguchi, H.~Zhang, and Y.~Mitsuya,
J. Chem. Phys. {\bf 121}, 4358 (2004).

\bibitem{harnau:04} L. Harnau, F. Penna, and S. Dietrich, Phys. Rev. E {\bf 70}, 021505 (2004).
\end{thebibliography}
\end{document}